\documentclass[showpacs,twocolumn,prl,floatfix]{revtex4}
\usepackage[dvips]{graphicx}
\usepackage[dvips]{color}
\newcommand{\href}[2]{#2}
\usepackage{amsmath}
\usepackage{amssymb}
\usepackage{bm}
\usepackage{times}

\newcommand{\kap}{\mbox{\boldmath $\kappa$}}

\begin{document}
\title{Fluctuation dissipation relations in stationary states of
interacting Brownian particles under shear}
\date{\today}
\author{Matthias Kr{\"u}ger and Matthias Fuchs}
\affiliation{Fachbereich Physik, Universit{\"a}t Konstanz, 78467 Konstanz, Germany}
\begin{abstract}
The fluctuation dissipation theorem (FDT) is studied close to the glass
transition in colloidal suspensions under steady shear. Shear breaks
detailed balance in the many-particle Smoluchowski equation, and
gives response functions in the stationary state which are smaller
at long times than estimated from the equilibrium FDT. During
the final shear-driven decay, an asymptotically constant relation
connects  response and fluctuations, restoring the form of the
FDT with, however, a ratio different from the
equilibrium one.
\end{abstract}


\maketitle In thermal equilibrium, the response of a system to a
small external perturbation follows directly from thermal
fluctuations of the unperturbed system. This connection is the
essence of the fluctuation dissipation theorem (FDT) which lies at
the heart of linear response theory. In non-equilibrium systems,
much work is devoted to understanding the general relation between
fluctuation (correlator $C(t)$) and response (susceptibility
$\chi(t)$) functions. It has been characterized by the fluctuation
dissipation ratio (FDR) $X(t)$ defined via
\begin{equation}
\chi(t) = - \frac{X(t)}{k_BT}\; \frac{\partial}{\partial t} C(t)\;
.\nonumber
\end{equation}
It is unity close to equilibrium, $X^{(e)}(t)\equiv1$, but deviates
in non-equilibrium because the external perturbations act against
non-vanishing currents (see Eq. \eqref{eq:genFDT} below); FDRs
quantify the currents and signal non-equilibrium \cite{unklar}.

Colloidal dispersions exhibit slow cooperative dynamics at high
concentrations and form glasses. These metastable soft solids can
easily be driven into stationary states far from equilibrium by
shearing with already modest flow rates. Spin glasses driven by
nonconservative forces were predicted to exhibit nontrivial
FDRs in mean field models \cite{Berthier99}. Such behavior was
observed in detailed computer simulations of sheared super-cooled
liquids by Berthier and Barrat \cite{Berthier02,Berthier02prl}.
During the shear induced relaxation, the FDR for particle motion
perpendicular to the shear plane is different from unity, but
constant in time. This ratio was also found to be independent of
observable, which led to the notion of an effective temperature
$X=T/T_{\rm eff}$ describing the non-equilibrium state. Further
simulations with shear also saw $T_{\rm eff}>T$ \cite{Ohern04,
Haxton07,Zamponi05}, and recently
$T_{\rm eff}$  was connected to
barrier crossing rates \cite{Ilg07}.

On the theoretical side, much effort has been made on different
spin-models, close to criticality. Universal FDRs  were found under coarsening \cite{Godreche00a}
and under shear \cite{Corberi03}, where at the critical temperature,
a universal value of $X=\frac{1}{2}$ was found.
Yet, the situation for structural glasses has not been clarified.

In this letter, we investigate FDT for colloidal suspensions close
to a glass transition under steady shear starting from the
$N$-particle Smoluchowski equation. While time dependent correlation
functions are calculated in the integration through transients (ITT)
approach  \cite{Fuchs02,Fuchs03}, which is based on mode coupling
theory, the connection to the corresponding susceptibilities will be
derived for the first time here. We show that equilibrium FDT is
violated, but can be restored in a well defined sense with a
renormalized FDR at long times; however, the ratio  depends on
variable, contradicting the notion of an effective temperature.
Moreover, we establish a connection to the concept of a yield
stress, which gives a scenario quite different from mean field spin
glass \cite{Berthier99}.

$N$ spherical Brownian particles of diameter $d$, with bare
diffusivity $D_0$, and interacting via internal forces ${\bf
F}_i=-\boldsymbol{\partial}_i U$, $i=1,\dots,N$, are dispersed in a
solvent
with a steady and homogeneous velocity profile ${\bf v}({\bf
r})=\kap\cdot{\bf r}$, with shear rate tensor
$\kap=\dot\gamma{\bf\hat x\hat y}$. Neglecting hydrodynamic
interactions, the distribution of particle positions evolves
according to the Smoluchowski equation \cite{Dhont}
\begin{equation}\label{smol}
\partial_t \Psi(t)=\Omega \; \Psi(t),\hspace{0.5cm} \Omega=\sum_{i}\boldsymbol{\partial}_i\cdot\left[\boldsymbol{\partial}_i-{\bf F}_i - \kap\cdot {\bf r}_i\right],
\end{equation}
where $\Omega$ is the Smoluchowski operator and we have introduced
dimensionless units for length, energy and time, $d=k_BT=D_0=1$. The
Smoluchowski operator for the system without shear ($\kap=\bf 0$)
and the flow-part will be denoted $\Omega_e$ and
$\delta\Omega=\Omega-\Omega_e$. We distinguish two time-independent
distributions, $\Omega_e \Psi_e=0$ without shear and
$\Omega\Psi_s=0$ for the stationary system. Averages are
$\left\langle\dots\right\rangle$ and
$\left\langle\dots\right\rangle^{(\dot\gamma)}$, respectively.
Stationary correlation functions are $C_{\rm ab}(t)=\langle \delta
a^*e^{\Omega^\dagger t}\delta b\rangle^{(\dot\gamma)}$, where
$\Omega^\dagger$ is the adjoint operator obtained by partial
integrations \cite{agarwal72,Fuchs02}; a fluctuation equals $\delta
a=a-\langle a\rangle^{(\dot\gamma)}$. Note that shear in
Eq.~(\ref{smol})
 leads to a non-Hermitian eigenvalue
problem \cite{graham}. The susceptibility $\chi_{\rm ab}(t)$ describes the linear response
of the stationary expectation value of $b$ to an external
perturbation $h_e(t)$ shifting the internal energy $U$ to
$U-a^*\,h_e(t)$,
\begin{equation*}
\left\langle b \right\rangle^{(\dot\gamma,h_e)}(t)-\left\langle b
\right\rangle^{(\dot\gamma)}=\int_{-\infty}^t \!\!\!\!\!dt'\chi_{\rm ab}(t-t')h_e(t')+\mathcal{O}(h_e^2)\; .
\end{equation*}
One finds $\chi_{\rm ab}(t)=\langle \sum_i \frac{\partial
a^*}{\partial {\bf r}_i}\cdot\boldsymbol{\partial}_i
e^{\Omega^\dagger t} b\rangle^{(\dot\gamma)}$ \cite{agarwal72}. In
non-equilibrium, where detailed balance is broken and a nonzero
stationary probability current ${\bf
j}_i^s=[-\boldsymbol{\partial}_i+{\bf F}_i + \kap\cdot {\bf
r}_i]\Psi_s$ exists, the equilibrium FDT is extended (with
$\boldsymbol{\hat\jmath}_i^\dagger$ the adjoint of the current
operator defined by ${\bf j}_i^s={\boldsymbol{\hat
\jmath}}_i\Psi_s$),
\begin{equation}
\Delta\chi_{\rm ab}(t)=\chi_{\rm ab}(t)+\dot C_{\rm ab}(t)=-\langle\sum\limits_{i}\boldsymbol{\hat\jmath}_i^\dagger\cdot\frac{\partial a^*}{\partial\mathbf{r}_i}e^{\Omega^{\dagger} t}b \rangle^{(\dot\gamma)}\label{eq:genFDT}
\end{equation}
and a deviation of the fluctuation dissipation ratio (FDR)
\begin{equation}
X_{\rm ab}(t)=\frac{\chi_{\rm ab}(t)}{-\dot C_{\rm ab}(t)}
\end{equation}
from unity, the value close to equilibrium, arises. While
Eq.~(\ref{eq:genFDT}) has been known since the work
of Agarwal \cite{agarwal72}, we will analyze it for driven
metastable (glassy) states and show that the additive  correction
$\Delta\chi_{\rm ab}(t)$ \cite{Blickle07, Speck06,Harada05} leads to
the nontrivial multiplicative correction, i.e., a constant FDR at
long times. For simplicity, we will look at
auto-correlations ($ b= a$) of $x$-independent fluctuations,
$\delta\Omega^\dagger a=0$, where the flow-term in the current
operator $\boldsymbol{\hat\jmath}_i^\dagger$ in (\ref{eq:genFDT})
vanishes.

$\Psi_s$ is not known and stationary averages are calculated via the
ITT approach \cite{Fuchs02,Fuchs03},
\begin{equation}
\langle \dots\rangle^{(\dot\gamma)}=\langle
\dots\rangle+\dot\gamma\int_0^\infty
ds\langle\sigma_{xy}e^{\Omega^\dagger s}\dots\rangle\; ,\nonumber
\end{equation}
where $\sigma_{xy}=-\sum_iF_i^x y_i$ is a microscopic stress tensor
element. (Operators act on everything to the right, except for when
marked differently by bracketing.) ITT simplifies the following
analysis because averages can now be evaluated in equilibrium, while
otherwise non-equilibrium forces would be required \cite{Szamel}.
E.g.~due to $\boldsymbol{\partial}_i \Psi_e={\bf F}_i \Psi_e$, the
expression (\ref{eq:genFDT}) vanishes in the equilibrium average.
The remaining term is split into three pieces containing
$\Omega^\dagger$
\begin{equation}
\Delta\chi_{\rm a}(t)=\frac{-\dot\gamma}{2}\int_0^\infty\!\!\!\!\! ds \langle
\sigma_{xy}e^{\Omega^\dagger s}
[
\textcolor{black}{\Omega^\dagger}  a^* -
 a^*\textcolor{black}{\Omega^\dagger} + (\Omega^\dagger a^*)]e^{\Omega^\dagger t}
  a\rangle\,.
\label{eq:dom}
\end{equation}
We start with the first term in the square brackets (without the factor $\frac{1}{2}$) which can be
integrated over $s$ directly giving
\begin{equation}
\dot\gamma\; \langle \sigma_{xy}  \delta a^*e^{\Omega^\dagger t} \delta a\rangle=
\left.\frac{\partial}{\partial t_w} C_{\rm a}(t,t_w)\right|_{t_w=0}\;,
\label{eq:chi}
\end{equation}
where from now on we consider  fluctuations from equilibrium $\delta
a=a-\langle a\rangle$; (the constant $\langle a\rangle$ cancels in
(\ref{eq:dom}) ). Intriguingly, in Eq.~(\ref{eq:chi})  the two time
correlator enters,
\begin{equation}
C_{\rm a}(t,t_w)=\langle  \delta a^{*}e^{\Omega^{\dagger} t} \delta a\rangle+\dot\gamma\!\int_0^{t_w} \!\!\!\!\!\!ds\langle\sigma_{xy}e^{\Omega^\dagger s} \delta a^{*}e^{\Omega^{\dagger} t}  \delta a \rangle,\label{eq:2ti}
\end{equation}
 where the rheometer has been shearing for a
period $t_w$ before the correlation measurement is started. It is
one of the central quantities in the spin-glass theory of aging
\cite{Berthier99}. While the transient correlator $C^{(t)}_{\rm
a}(t)=C_{\rm a}(t,0)=\langle  \delta a^{*}e^{\Omega^{\dagger} t}
\delta a\rangle$ describes the dynamics after switch on of the
rheometer, the stationary correlator $C_{\rm a}(t)=C_{\rm
a}(t,\infty)$ is observed after waiting long enough; it measures
fluctuations in the stationary state.

Our approximation for $\Delta\chi_{\rm a}(t)$ in
Eq.~(\ref{eq:genFDT}) rests on the observation that it contains the
product of a fluctuation $\delta a$ and the stationary current. We
expect current fluctuations to always decay to zero, even in
possible non-ergodic situations, and thus search for a coupling of
$\Delta\chi_{\rm a}(t)$ to derivatives of $C_{\rm a}(t)$ as they
cannot be non-ergodic. Partial integration can be used to show
$\frac{\partial}{\partial t_w} C_{\rm a}(t,t_w)|_{t_w=0}=\dot
C^{(t)}_{\rm a}(t)-\langle (\Omega_e^\dagger {a}^*)
e^{\Omega^\dagger t} \delta a\rangle$, where the latter term
contains the equilibrium derivative $\Omega^{\dagger}_e a^*$. It is
not conserved and decorrelates quickly as the particles loose memory
of their initial motion even without shear. The latter term then is
the time derivative of the equilibrium correlator, $C^{(e)}_{\rm
a}(t)=\langle \delta a^{*}e^{\Omega^{\dagger}_e t} \delta a\rangle$.
A shear flow switched on at $t=0$ should make the particles forget
their initial motion even faster, prompting us to use the
approximation $e^{\Omega^{\dagger} t} \approx  e^{\Omega^{\dagger}_e
t} P e^{-\Omega^{\dagger}_e t}\, e^{\Omega^{\dagger}t}$ with
projector $P=\delta a\rangle \langle \delta a^*   \delta a\rangle
^{-1}\langle \delta a^*$ in $\langle  (\Omega^\dagger_e
a^*)e^{\Omega^\dagger t} \delta a\rangle$; it is
then assured to decay faster than in equilibrium. This leads,
together with an analogous  approximation in $\langle \delta a^*
e^{\Omega^\dagger t} \delta a\rangle$, to
\begin{equation}
\left.\frac{\partial}{\partial t_w} C_{\rm a}(t,t_w)\right|_{t_w=0} \approx \dot C^{(t)}_{\rm a}(t)-\dot C^{(e)}_{\rm a}(t)\frac{C^{(t)}_{\rm a}(t)}{C^{(e)}_{\rm a}(t)}\label{eq:shti}.
\end{equation}
The last term in (\ref{eq:shti}) will be identified as {\it short
time derivative} of $C^{(t)}_{\rm a}$, connected with the shear
independent decay, while $\frac{\partial}{\partial t_w}
C_{\rm a}(t,t_w)|_{t_w=0}$ will turn out to be the {\it long time
derivative} of $C^{(t)}_{\rm a}$, connected with the final shear driven
decay. This is our main result. It captures the additional
dissipation provided by the coupling to the stationary probability
current in Eq.~(\ref{eq:genFDT}).
\begin{figure}
\begin{center}\includegraphics[width=0.967\linewidth]{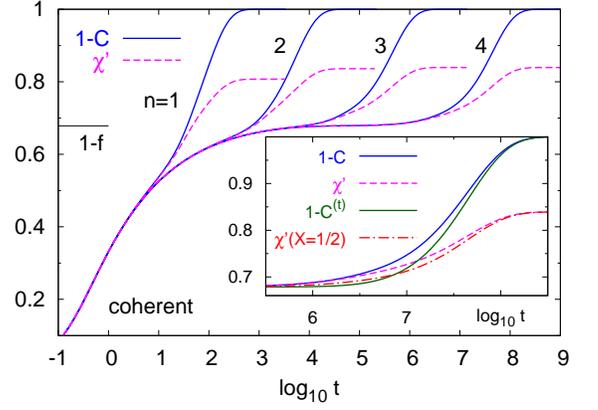}\end{center}
\caption{$C(t)$ from the $F_{12}^{(\dot\gamma)}$-model
\cite{supplement} and $\chi(t)$ via Eq.~(\ref{eq:FDT}) for a glassy
state ($\varepsilon=10^{-3}$) and $\dot\gamma=10^{-2n}$ with
$n=1...4$. Shown are integrated correlation, $1-C(t)$ and response
$\chi'(t)=\int_0^t\chi(t')dt'$. Inset shows additionally the
normalized transient correlator $C^{(t)}$ for comparison and the $\hat
X^{({\rm univ})}=\frac{1}{2}$ susceptibility for
$\dot\gamma=10^{-8}$.} \label{fig:schem1}
\end{figure}

In order to proceed, the difference between the stationary and the
transient correlators needs to be known. We will comment below on
the interesting result for the FDR following from the simplest
approximation to set them equal.  Going beyond this leading
approximation can be done via Eq. (\ref{eq:2ti})
\begin{equation}
C_{\rm a}(t)-C^{(t)}_{\rm a}\!(t)
\!\approx\!\int_0^\infty \!\!\!\!\!\!ds\frac{\langle\sigma_{xy}e^{\Omega^{\dagger} s}\sigma_{xy}\rangle}{\left\langle\sigma_{xy}\sigma_{xy}\right\rangle} \!\left.\frac{\partial C_{\rm a}(t,t_w)}{\partial t_w}\right|_{t_w=0},
\label{eq:stat1}
\end{equation}
where we used $t_w=\infty$, and factorized the appearing two-time
average with the projector
$\sigma_{xy}\rangle\langle\sigma_{xy}\sigma_{xy}\rangle^{-1}\langle\sigma_{xy}$.
A small parameter $\tilde\sigma\equiv\dot\gamma\int_0^\infty
ds\langle\sigma_{xy}\exp(\Omega^{\dagger}
s)\sigma_{xy}\rangle/\langle\sigma_{xy}\sigma_{xy}\rangle$ arises
which contains as numerator the stationary shear stress measured in
'flow curves' as function of shear rate \cite{Fuchs02}. For hard
spheres, the instantaneous shear modulus
$\langle\sigma_{xy}\sigma_{xy}\rangle$ diverges \cite{Dhont} giving
formally $\tilde\sigma=0$ and that transient and stationary
correlator agree. In recent simulations of density fluctuations of
soft spheres \cite{Zausch08}, the difference between the two
correlators
 was found to be largest at intermediate
times, and $C_{\rm a}(t)\leq C_{\rm a}^{(t)}(t)$ was observed. Both properties
are fulfilled by Eq.~(\ref{eq:stat1}).
\begin{figure}[t]
\begin{center}\includegraphics[width=0.967\linewidth]{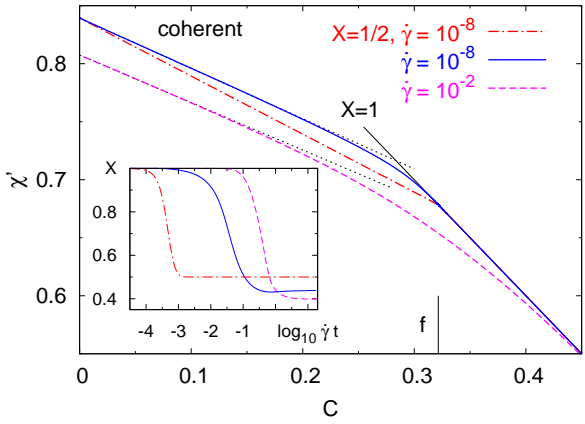}\end{center}
\caption{Parametric plot of correlation $C(t)$ versus response
$\chi'(t)=\int_0^t\chi(t')dt'$ for a glassy state
($\varepsilon=10^{-3}$) from the $F_{12}^{(\dot\gamma)}$-model
\cite{supplement} together with constant non-trivial FDR (straight
lines) at long times. The vertical solid line marks the plateau $f$.
Inset shows the FDR $X(t)$ as function of strain for the same
susceptibilities.} \label{fig:schem2}
\end{figure}

After the discussion of the first term in (\ref{eq:dom}), we turn to
the correction containing the last two terms in (\ref{eq:dom}). It
has vanishing initial value and in a mode coupling approximation in
ITT for the case of density fluctuations, the two terms also almost
cancel each other at long times making their sum a small correction.
Here, we proceed by ignoring it until a future presentation. We
hence find
\begin{equation}
\chi_{\rm a}(t)\approx-\dot C_{\rm a}(t)+\frac{1}{2}\left(\dot C^{(t)}_{\rm a}(t)-\dot C^{(e)}_{\rm a}(t)\frac{C^{(t)}_{\rm a}(t)}{C^{(e)}_{\rm a}(t)}\right).\label{eq:FDT}
\end{equation}
In the limit of small shear rates for glassy states, the correlators exhibit two separated
relaxation steps \cite{Fuchs03,Varnik06b}. During the shear
independent relaxation onto the plateau of height given by the
non-ergodicity parameter $f_{\rm a}$, we have $C^{(t)}_{\rm a}(t)\approx C^{(e)}_{\rm a}(t)$,
and the equilibrium FDT holds. During the shear-induced final
relaxation from $f_{\rm a}$ down to zero, i.e., for $\dot\gamma\to0$, and
$t\to\infty$ with $t \dot\gamma= $ const., the correlator without
shear stays on the plateau and its derivative is negligible. A
non-trivial FDR follows. Summarized we find in the glass
\begin{equation*}
\lim_{\dot\gamma\to 0} \chi_{\rm a}(t)=
\left\{ \begin{array}{ll}
-\dot C_{\rm a}(t) & \dot\gamma t\ll 1,\\
-\dot C_{\rm a}(t)+\frac{1}{2}\dot C^{(t)}_{\rm a}(t) & \dot\gamma t ={\cal O}(1).\\
\end{array}\right.
\end{equation*}
It is interesting to note that approximating stationary and
transient correlator to be equal \cite{Fuchs03}, $C^{(t)}_{\rm a}(t)\approx C_{\rm a}(t)$, we
find $\chi_{\rm a}(t)=-\frac{1}{2}\dot C_{\rm a}(t)$ for long times. The FDR in
this case takes the universal value
$\lim_{\dot\gamma\to0}X_{\rm a}(t\to\infty)=\hat X^{({\rm univ})}(\dot \gamma
t)=\frac{1}{2}$, independent of $a$. This is in
good agreement with the findings in \cite{Berthier02}. The initially
additive correction in Eq.~(\ref{eq:genFDT}) hence turns then into a
multiplicative one, which does not depend on rescaled time during
the complete final relaxation process.
\begin{figure}[t]
\begin{center}\includegraphics[width=0.967\linewidth]{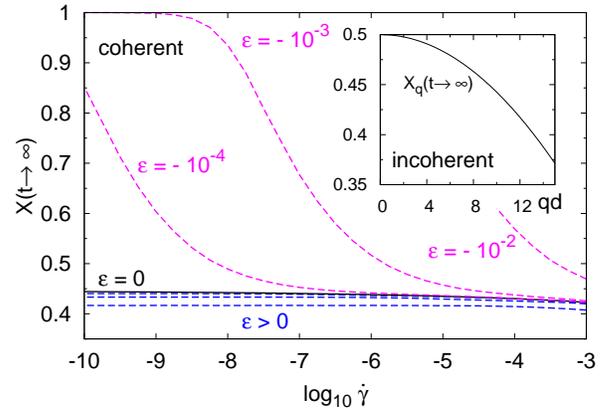}\end{center}
\caption{Long time FDR as function of shear rate for glassy
($\varepsilon\geq0$) and liquid ($\varepsilon<0$) states in the $F_{12}^{(\dot\gamma)}$-model \cite{supplement}, when approaching the glass transition for $\varepsilon=\pm10^{\{-2,-3,-4\}}$.
Inset: $\lim_{\dot\gamma\to 0} X_q(t\to\infty)$ as function of wavevector $q$ for incoherent density fluctuations at the critical density ($\varepsilon=0$) \cite{supplement}.
}
\label{fig:schem3}
\end{figure}

For a more precise investigation of the FDR, we have to consider the
difference between the transient and the stationary correlator in
Eq.~(\ref{eq:stat1}).  We  turn to the schematic
$F_{12}^{(\dot\gamma)}$-model of ITT \cite{supplement}, which has
repeatedly been used to investigate the dynamics of quiescent and
sheared dispersions \cite{Fuchs03}, and which provides excellent
fits to the flow curves from large scale simulations
\cite{Varnik06a}. It provides a normalized transient correlator
$C^{(t)}(t)$, as well as a quiescent one,
representing coherent, i.e., collective density fluctuations. The
corresponding stationary correlator $C$ is calculated in a second
step via Eq.~(\ref{eq:stat1}).
 Fig.~\ref{fig:schem1} shows the resulting
$\chi$ together with $C$ for a glassy state at different shear
rates. For short times, the equilibrium FDT is valid, while for long
times the susceptibility is smaller than expected from the
equilibrium FDT, this deviation is qualitatively similar for the
different shear rates. For the smallest shear rate, we also plot
$\chi$ calculated by Eq.~(\ref{eq:FDT}) with $\dot C_{\rm a}^{(t)}$
replaced by $\dot C_{\rm a}$, from which the universal $\hat
X^{({\rm univ})}(\dot\gamma t)=\frac{1}{2}$ follows. In the
parametric plot (Fig.~\ref{fig:schem2}), this leads to two perfect
lines with slopes $-1$ and $-\frac{1}{2}$ connected by a sharp kink
at the nonergodicity parameter $f$. For the other (realistic)
curves, this kink is smoothed out, but the long time part is still
well described by a straight line, i.e., the FDR is still almost
constant during the final relaxation process. We predict a
non-trivial time-independent FDR $\hat X_{\rm a}(\dot\gamma
t)=$const. if $C_{\rm a}^{(t)}$  (and with Eq. (\ref{eq:stat1}) also
$C_a$) decays exponentially for long times, because $\Delta\chi_{\rm
a}$ then decays exponentially with the same exponent. The line cuts
the FDT line {\it below} $f$ for $\dot\gamma\to 0$. All these
findings are in excellent agreement with the data in
\cite{Berthier02}. The FDR itself is of interest also, as function
of time (inset of Fig.~\ref{fig:schem2}). A rather sharp transition
from 1 to $\frac{1}{2}$ is observed when 
$C^{(t)}\approx C$ is approximated, which already takes
place at $\dot\gamma t\approx10^{-3}$, a time
 when the FDT violation is still invisible in Fig. \ref{fig:schem1}.
For the realistic curves, this transition happens two decades later. The huge difference is strikingly not apparent in the parametric
plot.
\begin{figure}[t]
\includegraphics[width=0.967\linewidth]{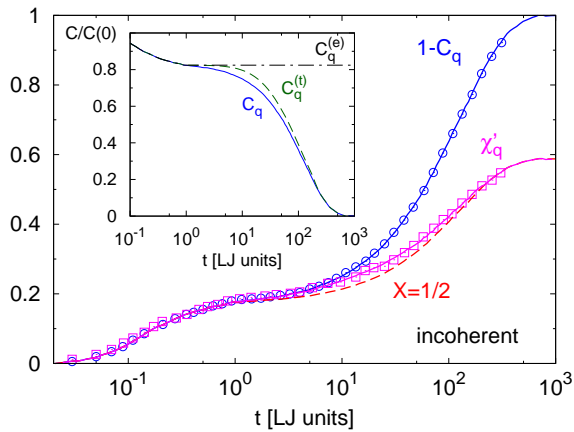}
\caption{\label{fig:fit}Comparison to simulation data for incoherent
density fluctuations in the neutral direction (wavevector ${\bf q}=7.47 {\bf
e}_z$) at temperature $T=0.3$ ($T_c=0.435$) and $\dot\gamma=10^{-3}$. Circles and squares are the data
(including units) from Fig. 11 in Ref. \cite{Berthier02}, lines are
$1-C_{{\bf q}}$ from Fig. 8 in Ref. \cite{Berthier02}, and the response
$\chi_{\bf q}'(t)=\int_0^t\chi_{\bf q}(t')dt'$ calculated via Eq. (\ref{eq:FDT}).  The dashed line shows $\chi_{\bf q}'$ with
approximation $C^{(t)}_{\bf q}\approx C_{\bf q}$. Inset shows the different correlators, see
main text.}
\end{figure}

Fig.~\ref{fig:schem3} shows the long time FDR as a function of shear
rate for different densities above and below the glass transition,
determined via fits to the parametric plot in the interval [0 :
0.1]. In the glass $X(t\to\infty)$ is nonanalytic while it goes to
unity in the fluid as $\dot\gamma\to0$, where we verified that the
FDT-violation starts quadratic in $\dot\gamma$ as is to be
expected due to symmetries. Our analysis also allows to study the
interesting question of the variable dependence of the
 FDR, for which we consider incoherent, i.e., single particle fluctuations \cite{supplement}
which were most extensively studied in
\cite{Berthier02}. The FDR is isotropic in the plane
perpendicular to the shear direction
 but not independent
of wavevector $q$, contradicting the idea of an effective
temperature as proposed in \cite{Berthier99,Berthier02} and others (see inset of Fig.~\ref{fig:schem3}).

That Eq.~(\ref{eq:FDT}) is nevertheless not in contradiction to the
data in Ref. \cite{Berthier02} can be seen by direct comparison to
their Fig. 11. For this, we need the quiescent as well as the
transient correlator as input. $C^{(e)}_{q}$ has been measured in
Ref. \cite{Varnik06b} suggesting that it can be approximated by a
straight line beginning on the plateau of $C_{{\bf q}}(t)$. In Fig.
\ref{fig:fit} we show the resulting susceptibilities. There is no
adjustable parameter, when $C^{(t)}_{\bf q}\approx C_{\bf q}$ is
taken, for the other curve, we calculated $C^{(t)}_{{\bf q}}(t)$
from Eq.~(\ref{eq:stat1}) using
 $\tilde\sigma=0.01$
(in LJ units) as fit parameter. The agreement is striking. In the
inset we show the original $C_{{\bf q}}$ from Ref.~\cite{Berthier02}
together with our construction of $C^{(e)}_{q}$ and the calculated
$C^{(t)}_{{\bf q}}$, which appears very reasonable comparing with
recent simulation data on $C_{\rm a}(t,t_w)$ \cite{Zausch08}.

In summary, shear flow drives metastable  Brownian dispersions to a
stationary non-equilibrium state with a multiplicative
renormalization of the FDR at long times, which is (almost)
independent of rescaled time.
It nearly agrees for variables not advected by flow
 and takes the universal value $\hat X_a(\dot\gamma
t)=\frac{1}{2}$ in glasses at small shear rates in leading
approximation. Corrections arise from the difference of the
stationary to the transient correlator, and depend on the considered
variable. They alter $\hat X_{\rm a}$ to values $\hat X_{\rm
a}\leq\frac{1}{2}$ in the glass. We show a new connection between
$\Delta\chi_{\rm a}$ and $C_{\rm a}(t,t_w)$, see Eq. \eqref{eq:chi},
which can be tested directly in simulations.

The derived FDRs characterize the shear-driven relaxation at long times,
which, according to the ITT approach, is also the origin of a
(dynamic) yield stress in shear molten glass  \cite{Fuchs02}. This
view captures extended simulations \cite{Berthier02,Varnik06a} and
broad-band experiments \cite{Sie:09}, establishing shear molten
glass as model for investigating non-equilibrium. Our finding of
values close to the universal $\hat X_a(\dot\gamma t)=\frac{1}{2}$
point to intriguing connections to critical systems
\cite{Godreche00a,Corberi03}. Open questions concern establishing
such a connection and to address the concept of an effective
temperature, which was developed for ageing and driven mean field
models.

We thank J.-L. Barrat, M. E. Cates and P. Ilg for helpful discussions.
M.K.~was supported by the DFG in IRTG 667.

\end{document}


\title{Fluctuation dissipation relations in stationary states of
interacting Brownian particles under shear: Supplementary material}
\author{Matthias Kr{\"u}ger and Matthias Fuchs}
\affiliation{Fachbereich Physik, Universit{\"a}t Konstanz, 78467 Konstanz, Germany}

\maketitle
This supplementary material summarizes the known ITT models for
transient correlators used to evaluate the stationary fluctuation
functions and susceptibilities presented in the main text.

\section{Schematic model}
In the schematic $F_{12}^{(\dot\gamma)}$-model of ITT, a
normalized transient correlator $\Phi(t)=C^{(t)}(t)$ is
considered which represents collective density fluctuations.
It obeys the equation of motion [12],
\begin{eqnarray*}
0&=&\dot\Phi(t)+\Phi(t)+\int_0^t dt'm(\dot\gamma,t-t')\dot\Phi(t'),\label{eq:f12}\\
m(\dot\gamma,t)&=&\frac{1}{1+(\dot\gamma
t)^2}\left[(v_1^c+2.41\varepsilon)\Phi(t)+v^c_2\Phi^2(t)\right],\label{eq:f122}
\end{eqnarray*}
with initialization $\Phi(t\to0)\to 1-t$. We use the much studied
values $v^c_2=2$, $v^c_1=v^{c}_2(\sqrt{4/v^{c}_2}-1)$ with the glass
transition at $\varepsilon=0$, and take $m(0,t)$ in order to
calculate quiescent ($\dot\gamma=0$) correlators \cite{Goetze84sup}. For the corresponding stationary correlator $C(t)$, we additionally have to estimate the normalized shear stress $\tilde\sigma$. In the $F_{12}^{(\dot\gamma)}$-model, the shear modulus is described by the transient correlator and we insert a factor of $\frac{1}{3}$ to account for the different plateau heights of the two normalized functions. $\tilde\sigma$ is hence estimated as
\begin{equation*}
\tilde\sigma=\dot\gamma\int_0^\infty \!\!dt\, \Phi(t)\frac{G_\infty}{f},
\end{equation*}
with $G_\infty/f\approx \frac{1}{3}$. Additionally we approximate the normalizations for $C^{(t)}(t)$ and $C(t)$, i.e., the respective structure factors, to be equal. For incoherent fluctuations, this is exact.

\section{Wavevector dependence}
To derive the long time FDR as function of wavevector $\bf q$ for incoherent, i.e., single particle density fluctuations, we use the normalized transient correlator from the ISHSM-model [12]. In glassy states, it approximately equals at long times
\begin{equation*}
C^{(t)}_{\bf q}(t)\approx f_q\exp\left(-\dot\gamma \frac{h_q}{f_q} t\right),
\end{equation*}
where $f_q$ is the $q$-dependent nonergodicity parameter and $h_q$ is another coefficient which as well as $f_q$ follows from equilibrium properties. The quiescent correlator is constant at long times, $C^{(e)}_{q}(t)=f_q$. The long time FDR then follows from Eqs. (3, 7-9) with $\tilde\sigma$ estimated as above.